\documentclass[lettersize,journal]{IEEEtran}
\usepackage{amsmath,amsfonts}
\usepackage{algorithmic}
\usepackage{algorithm}
\usepackage{array}
\usepackage[caption=false,font=normalsize,labelfont=sf,textfont=sf]{subfig}
\usepackage{textcomp}
\usepackage{stfloats}
\usepackage{url}
\usepackage{verbatim}
\usepackage{graphicx}
\usepackage{cite}
\usepackage{booktabs}
\usepackage{makecell}
\usepackage{color}
\usepackage{float}
\usepackage{subfig}

\begin{document}

\title{ActiveSelfHAR: Incorporating Self Training into Active Learning to Improve Cross-Subject Human Activity Recognition}

\author{Baichun Wei, Chunzhi Yi,~\IEEEmembership{Member,~IEEE,} Qi Zhang, Haiqi Zhu, Jianfei Zhu and \\Feng Jiang,~\IEEEmembership{Senior Member,~IEEE.}
\thanks{This work is funded by National Natural Science Foundation of China (No.62076080).}
\thanks{Baichun Wei and Chunzhi Yi share the same contribution. }
\thanks{Baichun Wei,Qi Zhang, Haiqi Zhu, Jianfei Zhu,and Feng Jiang are with the Faculty
of Computing, Harbin Institute of Technology, Harbin, Heilongjiang, China.}
\thanks{Chunzhi Yi and Feng Jiang are with the School of Medicine and Health,
Harbin Institute of Technology, Harbin, Heilongjiang, China.}
\thanks{Correspondence author: Feng Jiang (e-mail: fjiang@hit.edu.cn).}}

\markboth{Journal of \LaTeX\ Class Files,~Vol.~14, No.~8, August~2021}%
{Shell \MakeLowercase{\textit{et al.}}: A Sample Article Using IEEEtran.cls for IEEE Journals}

\IEEEpubid{0000--0000/00\$00.00~\copyright~2021 IEEE}

\maketitle

\begin{abstract}
Deep learning-based human activity recognition (HAR) methods have shown great promise in the applications of smart healthcare systems and wireless body sensor network (BSN). Despite their demonstrated performance in laboratory settings, the real-world implementation of such methods is still hindered by the cross-subject issue when adapting to new users. To solve this issue, we propose ActiveSelfHAR, a framework that combines active learning's benefit of sparsely acquiring data with actual labels and self-training's benefit of effectively utilizing unlabeled data to enable the deep model to adapt to the target domain, i.e., the new users. In this framework, the model trained in the last iteration or the source domain is firstly utilized to generate pseudo labels of the target-domain samples and construct a self-training set based on the confidence score. Second, we propose to use the spatio-temporal relationships among the samples in the non-self-training set to augment the core set selected by active learning. Finally, we combine the self-training set and the augmented core set to fine-tune the model. We demonstrate our method by comparing it with state-of-the-art methods on two IMU-based datasets and an EMG-based dataset. Our method presents similar HAR accuracies with the upper bound, i.e. fully supervised fine-tuning with less than 1\% labeled data of the target dataset and significantly improves data efficiency and time cost. Our work highlights the potential of implementing user-independent HAR methods into smart healthcare systems and BSN.
\end{abstract}


\section{Introduction}
\IEEEPARstart{W}{ith} the development of the internet of things (IoT) and smart infrastructure, HAR has become an indispensable technology for emerging smart environment applications, such as augmented reality, participatory sensing, and health IoTs \cite{chettri2019comprehensive,gao2015survey,chadwell2020technology,aceto2020industry}. The growing prevalence of smart wearable devices improves the energy cost and site constraints of high-resolution cameras and radars, thus enabling pervasive usages of HAR in daily application scenarios \cite{kim2019wearable,porciuncula2018wearable,zhang2022deep}. Although recent algorithmic advancements demonstrate their performance with multiple open-sourced datasets, the real-world implementation of such HAR methods is still hindered by the cross-subject issue of applying such methods to new users \cite{soleimani2021cross,zhang2020dual,li2019adaptive}.

The cross-subject issue denotes a significant decline in activity recognition accuracy in new-user implementation due to the wearable sensor signals’ sensitivity to sensor placement and user-specific characteristics \cite{fallahzadeh2017personalization}. Although deep learning (DL) techniques present the ability to automatically learn the representation of data distributions and proved effectiveness over traditional hand-crafted features, the issue is still problematic \cite{bird2020cross}. When a HAR classifier needs to be adapted to new users, an intuitive method is supervised transfer learning (e.g., the most commonly used fine-tuning framework) to reduce the data collecting and labeling burden \cite{abdel2020st}. The fine-tuning framework fixes the parameters of the shared network and uses a medium number of labeled samples of the target subject or dataset to fine-tune the parameters of the learned network \cite{girshick2014rich, perera2019learning}. This process requires lower time and labor costs than fully supervised learning. However, unlike the interpretable data of computer vision and natural language processing, the sensor data of HAR is hard to label without the reliance on specific equipment, such as the foot pressure sensor. Thus, it is challenging to construct a medium-sized dataset of the new user to satisfy the need for model fine-tuning \cite{hu2018fusion}.

Alternatively, unsupervised domain adaptation (UDA) has been adopted to solve the cross-subject issue without data labeling \cite{wang2018deep,ding2018empirical,fallahzadeh2021personalized}. In this framework, the large dataset for model pre-training is referred to as the source domain, and the target subject or dataset is referred to as the target domain \cite{bousmalis2016domain}. The UDA leverages the difference in intrinsic structure of the source and target domain to obtain the supervision information, such as the domain adversarial methods \cite{ganin2016domain}. Through this process, the HAR model is updated and implemented to classify the data of the target domain. Despite the proven feasibility of UDA, there still exist some limitations. First, the supervision information of UDA is easily biased by the difference in the source and target domain, which makes it less reliable than the actual labels \cite{reed2014training}. Second, in the training process of UDA, the model not only needs to learn to differentiate among classes, but also needs to extract the supervision information based on the source and target domain. The increased tasks would cause the increased time and computation cost in the model training process \cite{tzeng2017adversarial}. 

Semi-supervised learning frameworks ease the conflict between the relatively large data labeling cost and the limited performance improvement of unsupervised methods using sparsely or noisily labeled data, such as active learning \cite{kumar2020active}. Active learning selects the most informative samples from the target domain to query their labels and to construct a core set that covers diverse labels and data distributions of samples \cite{hossain2015sleep}. In this way, the HAR model can be trained over the core set and learns the distribution of the target domain. Although multiple strategies have been proposed to select informative samples and train the model, the DL's requirement of data quantity remains the conflict between the amount of labeled data and model performance \cite{sener2017active}.

As another semi-supervised technique, self-training methods leverage the large-scale unlabeled data based on an iterative training process \cite{zoph2020rethinking}. In self-training framework, a teacher model is first initialized with a small target labeled dataset. Then, the teacher model is utilized to generate self-training labels for the unlabeled data pool and select the samples with high confidence to augment the labeled dataset. At last, a student model is trained or fine-tuned based on the augmented dataset. Such methods demonstrate the feasibility in the HAR field \cite{saeed2019multi}. Practically, pre-training a deep model with a small dataset may lead to model overfitting or underfitting, either of which may increase the noise of the pseudo labels and even may induce a worsened performance.

When we consider the requirements of real-world scenarios, there is a research gap between currently developed methods and the supervision information we can obtain. When adapting to a new user (e.g., a smartphone or a smartwatch), it is commonly user-acceptable to have some short-time calibrations (i.e., sparse queries) to improve the performance with the continuous usage of the algorithm (e.g., iterative training). Thus, in this study, we propose to combine active learning's benefit of sparsely acquiring data with actual labels and self-training's benefit of effectively utilizing unlabeled data. We expect to use the iterative training manner to reduce the amount of queried data and use the sparsely queried labels to accelerate the convergence of self-training and improve the overall performance.  Particularly, we first select the high confidence samples as the self-training set according to the soft labels predicted by the pre-trained model. Second, we evaluate the distance between the samples of the non-self-training set and the self-training set, then select the top list as the core set for querying. In this way, the samples with the lowest confidence can be selected as informative samples and queried for labels. Third, we cluster the samples of the non-self-training set through the spatio-temporal relationships and assign the queried labels to the samples, thus augmenting the core set accordingly. Finally, the model is finetuned by the training set constructed by the self-training and augmented core set and prepared for the next iteration. Our contributions can be summarized as follows:
\begin{itemize}
\item To the best of our knowledge, our proposed ActiveSelfHAR is the first framework that complements active learning with the self-training framework and can effectively select and utilize the sparsely queried samples.
\item We propose to utilize the spatio-temporal relationship of low-confidence samples to cluster and label the non-self-training set samples to improve the diversity and quantity of the core set. In this way, both the distribution similarity and HAR's nature of time continuity can be considered.
\item We evaluate our proposed framework on various EMG and IMU-based HAR datasets. The proposed approach shows similar results to the upper bound method, i.e., supervised fine-tuning framework.
\end{itemize}

\section{Related Work}
\subsection{Human Activity Recognition}
The HAR task has been studied for years, typically detailed as recognizing the locomotion mode of a user (e.g., level-walking, stair ascend,and ramp descend) based on the data of the user and the surroundings. Based on the type of instrumented devices, the HAR techniques can be divided into two categories: sensor-based HAR and video-based HAR \cite{ranasinghe2016review}. Herein, we only introduce the sensor-based HAR method to match our research scope. 

In recent years, deep learning methods, such as convolutional neural networks (CNN), have become more prevalent in HAR tasks \cite{atzori2016deep}. The implementation of CNN has enabled the joint training of the feature extractor and the classifier, making the features more task-oriented than traditional hand-crafted features. Saeed et al. proposed the transformation prediction network (TPN), which simultaneously learned to solve the signal transformation and activity recognition tasks based on temporal convolution filters \cite{saeed2019multi}. Su et al. introduced a hierarchical CNN architecture for the HAR task, which learned the level-based features from the sensor signals without prior knowledge \cite{su2019cnn}. 

Although the HAR algorithms demonstrate their performance on the open-sourced dataset, the real-world application is still hindered by the covariate shift when generalizing the algorithms across different users. That is, the algorithms' performance usually suffers from a significant degradation when applied to new users due to the user-dependent characteristics of wearable sensors \cite{cote2019deep}. The collection of labeled data that is sufficient to train a new user-specific model is time and labor-expensive. It is of vital importance to ease the issue from an algorithmic perspective.

\subsection{Transfer Learning}
Transfer Learning is a commonly used method to mitigate the data collecting limitation in the cross-subject problem. Suppose correlations exist between the large source dataset and the target user. In that case, transfer learning enables the model to adapt to the user by applying the knowledge learned from the source dataset \cite{tan2018survey}.  

Supervised transfer learning is a practical framework for model transfer based on a middle-sized labeled dataset. The most commonly used method is fine-tuning, where a deep network is first pre-trained on a large dataset, and then part of the network is fine-tuned by the labeled data of a different domain. In our previous study, we implement the fine-tuning method to transfer our model from a normal walking-based HAR task to an exoskeleton assisting-based HAR task, increasing the accuracy from 72.63$\%$ to 93.61$\%$ \cite{wei2021novel}. However, labeling a moderate dataset would still be a heavy burden since it is difficult to label the sensor data visually. 

Since the data labeling of HAR is burdensome for both the user and the researcher, some researchers leverage unsupervised domain adaptation to address the cross-subject problem. In the UDA framework, the pre-trained model is updated based on the supervision information extracted from the domain intrinsic structures. Fallahzadeh et al. proposed a cluster-based cross-subject UDA method, which leverages the similarities between the users to extract the core observations in the target domain \cite{fallahzadeh2017personalization}. Zhang et al. leveraged the domain-adversarial techniques to build a UDA framework for cross-subject HAR, which updated the feature generator and classifier step by step based on maximizing the classification discrepancy between the source domain and the target domain \cite{zhang2020unsupervised}. Although the UDA techniques have reduced the burden of data labeling, they have introduced more computational costs in the model training phase \cite{chen2016training}. In addition, the supervised information of UDA is extracted based on the domain correlation, which is less reliable and intuitive than the actual labels \cite{reed2014training}.

\subsection{Semi-supervised Learning}
Given the limitations of the supervised and unsupervised transfer learning techniques, some researchers introduce model adaptation methods based on semi-supervised learning. Semi-supervised learning aims to utilize unlabeled data to complement the sparsely labeled data to improve the quantity and diversity of the target dataset can be improved. Saeed et al. proposed the SelfHAR: a model adaptation method that leverages the self-training framework \cite{tang2021selfhar}. Also, they reinforced the feature extractor of the teacher model based on a large unlabeled dataset and the self-supervised learning framework. The teacher model's feature extractor was shared with the student model during student model fine-tuning. Although this method has been proven effective, it still needs a labeled dataset with a certain scale (about 10$\%$ compared with fully supervised learning) so that the teacher model would be able to construct the self-training dataset. Moreover, the self-training framework ignores the requirement of actively selecting the to-be-labeled samples. Thus, it would suffer from an inefficient labeled dataset and consequently degraded performance.

Active learning has been used to complement the HAR systems in recent years, aiming to attaining labels at a low-cost \cite{hossain2018deactive}. Instead of a dense data labeling strategy for the whole dataset, the goal of active learning is to sparsely select the most informative data point to label so that the model trained over the small core set is competitive over the whole data \cite{sener2017active}. Alemdar et al. proposed an active learning-based HAR system, which could reduce the labeling cost by 30$\%$-75$\%$ \cite{alemdar2017active}. For an active learning-based user-independent HAR system, one of the main concerns is the balance between the budget of data labeling and the scale of the core set for model training. In order to augment the core set without the need for additional active labeling, Hossain et al. introduced a cluster-based data labeling strategy to assign the label of the most informative sample of each cluster to the samples in the same cluster \cite{hossain2017active}. This method has enriched the core set through the spatial relationship between samples. However, other prior knowledge in human activity, such as the temporal relationship, can be leveraged to increase the diversity and quantity of the core set. 

Furthermore, the active learning-based methods to some extent ignore the usage of the source domain model and the potential benefit of iterative training paradigm. In our study, we combine the iterative training paradigm, the source-domain knowledge from the self-training framework, and the active sample selection benefit of active learning. In so doing, we reduce the amount of queried data and thus improve the overall performance.

\section{Method}
\subsection{Problem Definition}
In this section, we formally defined the problem of cross-subject HAR. According to \cite{sener2017active}, the active learning loss $E_{AL}$ is upper bounded by the summation of the generalization error $E_{G}$, training error $E_{T}$, and core set loss $E_{C}$. When the classifier is specific to CNN, reducing the core-set loss is critical for optimizing the total loss. 
\begin{equation}
\label{eq_01}
E_{AL} = E_{x,y \sim pz}[l(x,y; A_{s})] \leq E_{G} + E_{T} + E_{C}
\end{equation}

\begin{equation}
\begin{split}
\label{eq_02}
E_{G} &= | E_{x,y \sim pz}[l(x,y; A_{s})] - \frac{1}{n} \sum_{i \in [n]} l(x_{i},y_{i}; A_{s}) | \\
E_{T} &= \frac{1}{|s|} \sum_{i \in [n]} l(x_{j},y_{j}; A_{s}) \\
E_{C} &= |\frac{1}{n} \sum_{i \in [n]} l(x_{i},y_{i}; A_{s}) - \frac{1}{|s|} \sum_{j \in [n]} l(x_{j},y_{j}; A_{s})| \\
\end{split}
\end{equation}

Compared to approximating the core set loss as solving the K-center problem that minimizes the distance between each data point and its nearest center, we herein formulate the problem in a task-specific manner. Particularly, we consider the requirement of sparse query and the refinement benefit of self-training. Thus, we combine active learning with probabilistic labels to screen informative points whose classifications the classifier is not confident with. In this way, we can query labels sparsely and effectively. We formulate the problem as:

\begin{figure*}[t]
\centering
\subfloat[]{\includegraphics[width=17cm]{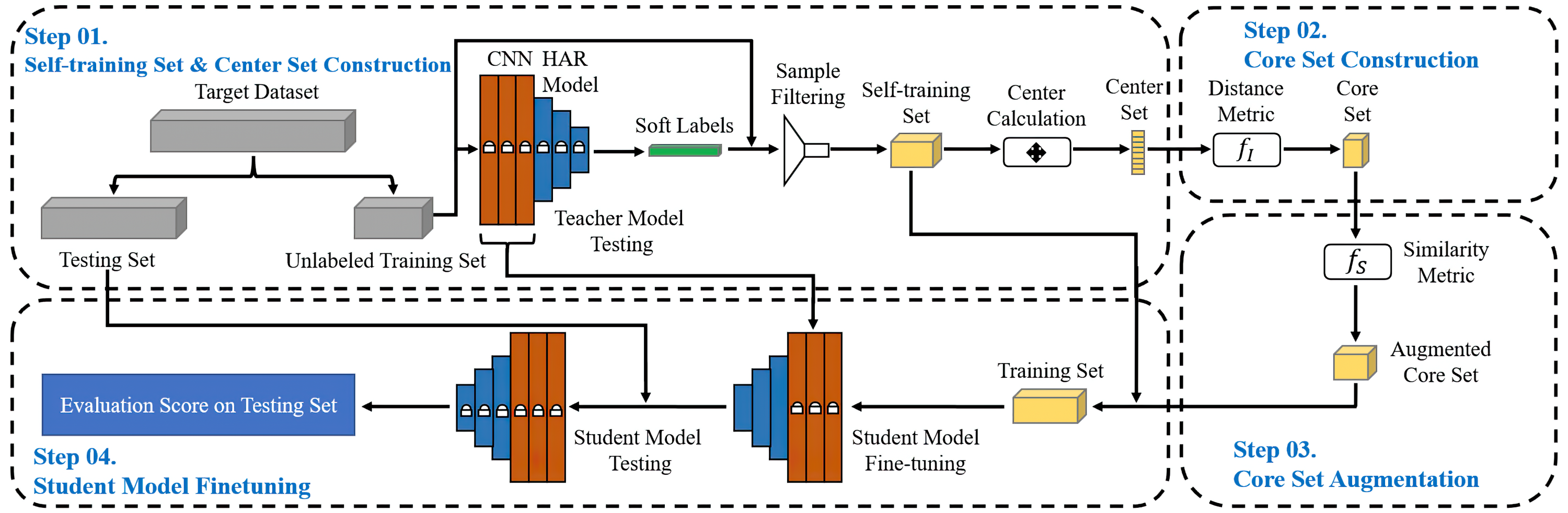}%
\label{fig1_01}}
\hfil
\subfloat[]{\includegraphics[width=17cm]{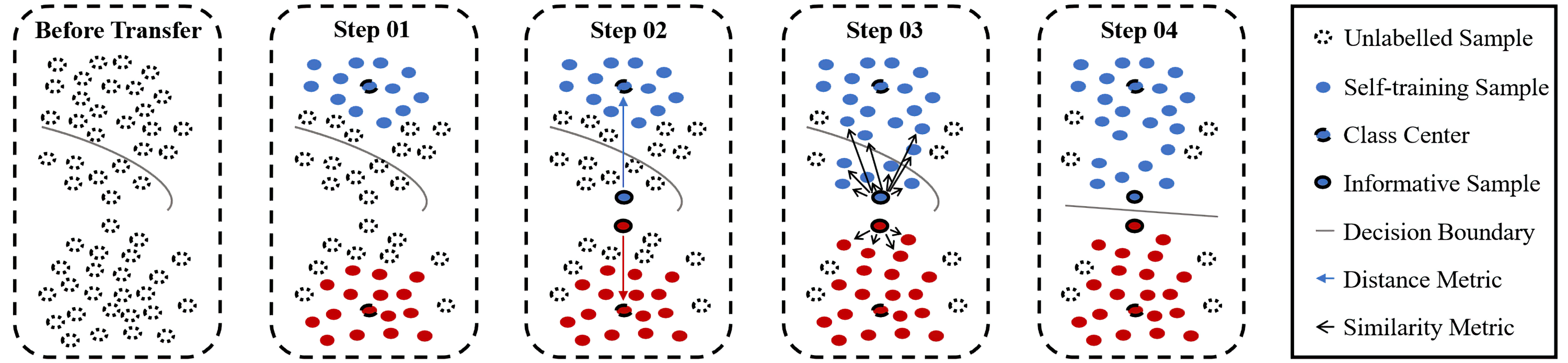}%
\label{fig1_02}}
\caption{Overview of the proposed ActiveSelfHAR framework.(a) The detailed schematic of the data labeling and model training pipeline that consists of four steps. (b) A visual explanation of how to construct a labeled subset from the unlabeled target dataset in four steps of the framework.}
\label{fig_1}
\vspace{-10pt}
\end{figure*}

\begin{equation}
\begin{split}
\label{eq_05}
I &= \mathop{argmin}\limits_{i \in T\backslash[S]} (I_{aug}+ \tilde{y}_{i}^{s} + \Omega (I_{Q})), j\sim T\backslash[Q, i] \\
I_{aug} &= \mathop{max}\limits_{j}(sim(\tilde{x}_{i}^{q},{x}_{j}) + cont(i,j))
\end{split}
\end{equation}
\begin{equation}
\label{eq_06}
\theta^{*} = argmin\sum_{k\in I\cup S\cup J}L(x_{k},y_{k}; \theta)
\end{equation}
where $\tilde{y}_{i}^{s}$ denotes the probabilistic labels or confidence scores predicted by the pre-trained source model or self-trained model of the last iteration, $\tilde{x}_{i}^{q}$ denotes the most informative samples (i.e., the core set $Q$), $x_{j}$ denotes the low-confidence samples except for the core set, $T$ denotes the target domain, $sim$ denotes the similarity between two data points, $cont$ denotes the temporal continuity of the same cluster, $I_{aug}$ denotes the index set of the augmented core set expect for the core set $Q$, $I_{Q}$ denotes the index set of the core set, $\Omega(I_{Q})$ denotes the dimension of the index set, $S$ denotes the self-training set, $J$ denotes the index set of the low-confidence samples consisting of $x_{j}$, $L(x_{k},y_{k};\theta)$ denotes the loss of the prediction model parameterized by $\theta$. Equation \eqref{eq_05} selects data points with low confidence scores, and determines the informative points $\tilde{x}_{i}^{q}$ from the selected samples. Then, the actual labels of data center points are queried and assigned to the points $x_{j}$ in the same clusters to construct the augmented core set. Equation \eqref{eq_06} combine the self-training set and the augmented core set to to obtain new parameters $\theta^{*}$. The whole progress can be formulated as an Expectation-Maximum (EM) issue, given by: 

\begin{equation}
\label{eq_07}
\theta^{k+1} = argmin\sum logP(S,I|\theta^{k})\cdot P(I|S,\theta^{k})
\end{equation}
The issue can thus be solved in an EM manner. 

\subsection{Framework of ActiveSelfHAR}
\subsubsection{Overview} The proposed framework combines active learning and self-training framework into a multi-step model training pipeline (Fig. 1):

1. We employ the model trained in the last iteration as the teacher model to select a self-training set $S$ from the unlabeled training set $T$ based on a confidence threshold. After that, the center $c_{k}$ of class $k$ in the set is determined by calculating the centroid of all data in that class, We refer to the set of all $c_{k}$ as the center set $C$. 

2. For each $x_{i}\in T\backslash S$, we calculate its Euclidean distance to each center in $C$ to form a distance vector $v_{i}\in R^{K}$, where K denotes total number of classes. We further determine the most informative samples to query for labels and form a core set $Q$ based on their distance vectors and a distance metric function $f_{I}$. It should be noted that we assume that the labels queried by active learning are actual labels and can be obtained online.

3. For each $x_{i}^{q}\in Q$, we first creat a cluster that contains samples in the temporal neighborhood of $x_{i}^{q}$, and then assign the label of $x_{i}^{q}$ to its similar samples in the cluster based on the spatial relationships.The temporal and spatial relationships are evaluated based on a similarity metric function $f_{S}$. Then we combine the core set and the newly labeled samples to form a augmented core set $A$.  

4. The self-training set S and augmented core set A are combined to form a labeled training set $T^{'}$ for student model training. During the model training process, the parameters of the teacher model's CNN layers are shared to the student model of the same network architecture, while the remaining layers of the student model are fine-tuned based on the training set $T^{'}$.  

\subsubsection{Self-training Set and Center Set Construction} In this step, we construct the self-training set $S$ based on the self-training framework. First, the model trained in the last iteration is employed as the teacher model. Then, we use the teacher model to classify the unlabeled training set and treat the outputs of the softmax layer as the probabilistic labels. In addition to the label generation, we extract the outputs of the penultimate layer as the features of the training samples, and reduce the feature dimension to three-dimension using principal component analysis (PCA). This process enables the subsequent data analysis process to be carried out in a low-dimensional space since the Euclidean distance used in this paper is less efficient in the high-dimensional space. 

After probabilistic label generation and feature dimension reduction, we set a confidence threshold to select the samples with the score (the max value of the corresponding probabilistic label) greater than the threshold to construct the core set. In this paper, the confidence thresholds for the EMG-based and IMU-based HAR dataset are 0.5 and 0.3, respectively. We set a higher threshold for the EMG-based dataset since it contains fewer classes. For each iteration, the threshold increases by 0.05 to reduce the noise of the core set labels since we assume that the model is more competitive as the number of iterations increases. After self-training set construction, we calculate the data centroid of each class in the set. Then, for each class $k$ of the set, we take the nearest sample to the class centroid as the center $c_{k}$ of the class. At last, the center set $C$ is constructed with the class centers.

\subsubsection{Core Set Construction} For each three-dimensional sample $x_{i} \in T\backslash S $, we calculate its Euclidean distance to each $c_{k}$. Then, we evaluate the information carried by each sample through a distance-based metric:
\begin{equation}
\label{eq_08}
f_{I}(x_{i}) = \frac{d(x_{i},c_{sn})-d(x_{i},c_{n})}{d(x_{i},c_{sn})}
\end{equation}
where $d(x_{i},c_{sn})$ and $d(x_{i},c_{n})$ denote the distance from the $x_{i}$ to the second nearest and nearest center, respectively. From \eqref{eq_08}, it is evident that $0 < f_{I} (x_{i}) < 1$. The closer $f_{I} (x_{i})$ is to 0, the lower confidence $x_{i}$  has, and thus the more informative $x_{i}$ is. 

To mitigate the biases of data labeling, we pay attention not only to the information carried by the samples but also to the spatial location of the samples. We creat $K\cdot (K-1)/2$ categories corresponding to $K\cdot (K-1)/2$ class boundaries, where K represents the total number of the classes. Then, we match each sample with a category based on its nearest and second nearest centers. After that, for each category, we select top $N$ samples with the highest $f_{I}$ to query for label, resulting in $N\cdot K\cdot(K-1)/2$ labeled data to construct a core set $Q$. In this paper, we set the number $N$ to 10.

\subsubsection{Core Set Augmentation} Once the core set $Q$ is formed, we leverage each $x_{i}^{q}\in Q$ to select the less informative samples that are likely in the same class as $x_{i}^{q}$ based on the proposed spatio-temporal similarity metric: 
\begin{equation}
\label{eq_09}
f_{S}(x_{j}) = \frac{d(x_{j},c_{in})}{d(x_{i}^{q},c_{in})}+\frac{abs(ts_{x_{i}^{q}}-ts_{x_{j}})}{thres_{t}}
\end{equation}
where $d(x_{i}^{q},c_{in})$ denotes the distance from $x_{i}^{q}$ to the center of the same class, $d(x_{j},c_{in})$ denotes the distance from sample $x_{j}$ to the same center, $abs()$ denotes the absolute operation, $ts_{x_{j}}$ denotes the timestamp of the sample $x_{j}$, and $thres_{t}$ is a threshold to measure the temporal relationship between $x_{i}^{q}$ and $x_{j}$. In this paper, we set the $thres_{t}$ to 5 s and only select the samples with $f_{S}(x_{j})\leq 1$ to be assigned with the label of $x_{i}^{q}$. In this way, the samples of high temporal correlations with $x_{i}^{q}$ and of less informative than $x_{i}^{q}$ are selected and labeled since the temporal-related samples that are closer to the center are more likely to in the same class with $x_{i}^{q}$. Then, these newly labeled samples are used to constructed an augmented set $A$ with the core set $Q$. 

\subsubsection{Student Model Fine-tuning} In this step, we first construct a training set $T^{'}$ based on the augmented set $A$ and self-training set $S$. Since the teacher model has been heavily trained through the source data and the target data in the previous iterations, we adopt the fine-tuning framework in consideration of  computational cost efficiency. In this framework, the CNN layers of the teacher model are shared with the student model, while the fully-connected layers of the student model are updated based on the training set $T^{'}$.

\subsection{Configurations of ActiveSelfHAR}
\subsubsection{Data Processing} In this paper, we adopted a sliding window scheme to divide the data sequence into  segments \cite{englehart2003robust}. The timeste of each analysis window is set to 300 ms, which is in the suitable range of window length for HAR tasks (200 ms - 500 ms). The timestep between the adjacent windows is set to 30 ms so that a dense dataset can be generated. 

\subsubsection{CNN for Human Activity Recognition} CNN has been widely implemented in HAR tasks, which is able to capture the spatio-temporal relationship between multi-channel signals in a sparse manner. For the EMG-based HAR dataset, we adopt the transformation prediction network (TPN), which leverages the 1-D convolutional layers to extract features along the temporal dimension \cite{saeed2019multi}. The architecture of TPN is shown in Fig. 2(a). The three convolutional layers consist of 32, 64, and 96 feature maps with kernel sizes of 24, 16, and 8 ,respectively. The stride of the kernels is 1. After each layer, drop out with a rate of 0.1 and L2 regularization with a rate of 0.0001 is used. A global 1D maximum pooling layer is connected to the last convolutional layer to serve as the output layer of CNN architecture. The original fully-connected architecture of TPN contains two layers, a transition layer with 1024 units and an output layer with the number of units equal to the classes. Since the outputs of the penultimate layer are treated as the features and are further reduced to 3-D, we add a fully-connected layer with 32 units between the transform layer and the output layer. ReLU activation function is applied after each layer except for the output layer.

For the IMU-based HAR dataset, the analysis window length is 300 ms, and the data sampling frequency is 100 Hz, resulting in 30 sample points in each window. Such a small window size cannot enable the analysis window to go through the TPN network. Thus, given the recent satisfactory performance of the multi-stream CNN architectures, we construct a double-stream CNN for the IMU-based HAR task \cite{su2019cnn,wei2019surface}. As shown in Fig. 2(b), each CNN stream contains three layers with 32, 64, and 64 feature maps, respectively. The first convolutional layer with a kernel size of 3×1 and stride of 1 is used to extract the local temporal features of each channel. The second convolutional layer with kernel size of 6×3 and stride of 3 is used to reduce the temporal and spatial dimension of each sensor since the data of IMU is 3-axis. The third convolutional layer with a kernel size of 6×1 and stride of 1 is used to extract the features of each channel with a broader temporal receptive field, and the layer with a kernel size of 3×3 and stride of 1 is used to extract the spatial relationship between the channels. ReLU activation function and batch normalization are applied after each layer except for the output layer. 

\begin{figure}[h]
\centering
\subfloat[]{\includegraphics[width=9cm]{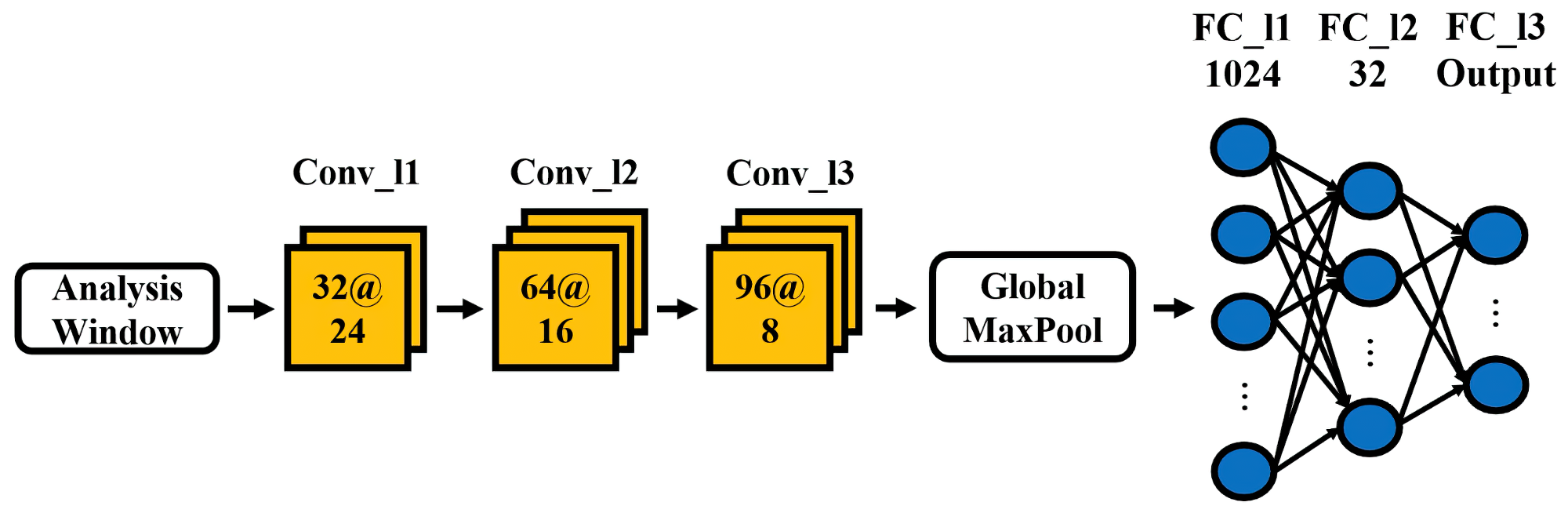}%
\label{fig2_01}}
\hfil
\subfloat[]{\includegraphics[width=9cm]{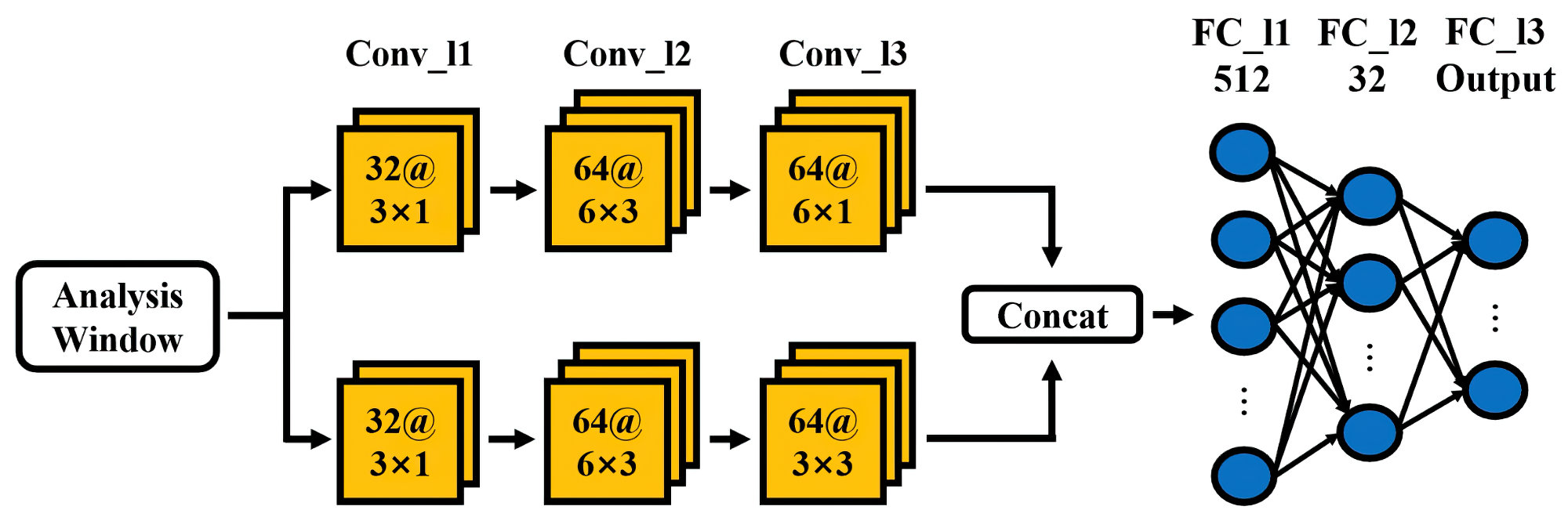}%
\label{fig2_02}}
\caption{The architecture of (a) TPN network and (b) the proposed double-stream CNN.}
\label{fig_2}
\end{figure}

\subsubsection{Training and Testing Setup} Since we focus on the cross-subject problem in the HAR task, we design a modified leave-one-out cross-validation process. Each time, a subject is selected to be the target subject, and the HAR model is pre-trained based on the rest of the subjects. After that, the data of the target subject is split into training and testing sets with a ratio of 3:4. The training set is used for model transfer based on the ActiveSelfHAR. The learning rate and the batch size of DSAD, PAMAP, and our datasets (described in Section IV-A) are 0.00001/512, 0.0000001/256, and 0.000001/64, respectively. Adam optimizer is used to train the network with cross-entropy loss function. Additionally, since active learning is an iterative model adaptation process, we set the maximum number of iterations of our method to be 3. In each iteration, the number of epochs is set to 30, which is a commonly used number in HAR methods.
\begin{equation}
\label{eq_10}
loss(x,y) = E_{(x,y)\in(X,Y)}\sum_{n=1}^{N}-I[n=y]logP_{n}(y|x)
\end{equation}
where $E$ is the expectation operator, $I$ is the binary indicator, and $P_{n}$ is the output probability of class n. After the model transfer process, the testing set is used for model evaluation.

\section{Experiment Protocols}
\subsection{Experimental Datasets} In this paper, considering the commonly used electromyography (EMG) and inertial measurement unit (IMU) signals for HAR, we use three datasets to evaluate the performance of our method. Two of them are IMU-based datasets, and one of them is an EMG-based dataset. The sensor placement in each dataset is shown in Fig. 3. A description of each dataset is provided below.

One of the IMU-based datasets is the daily and sports activities dataset (DSADS), which contains 8 subjects performing a large amount of activies (19 activties) \cite{barshan2014recognizing}. For each activity, the subjects performed at least five minutes. Five 9-axis IMUs with a frequency of 25 Hz are mounted on the subjects. We chose 12 most commonly occurred activities in daily time to evaluate our method: sitting, standing, lying on the back, lying on the right side, ascending stairs, descending stairs, walking, ascending ramp, running, stepper exercise, cross trainer exercise, and jumping. Additionally, the sampling frequency of IMUs is resampled to 100 Hz to generate a denser dataset and to make the sampling frequency equal to another IMU-based dataset.

\begin{figure}[bhtp]
  \centering
  \includegraphics[width=\linewidth]{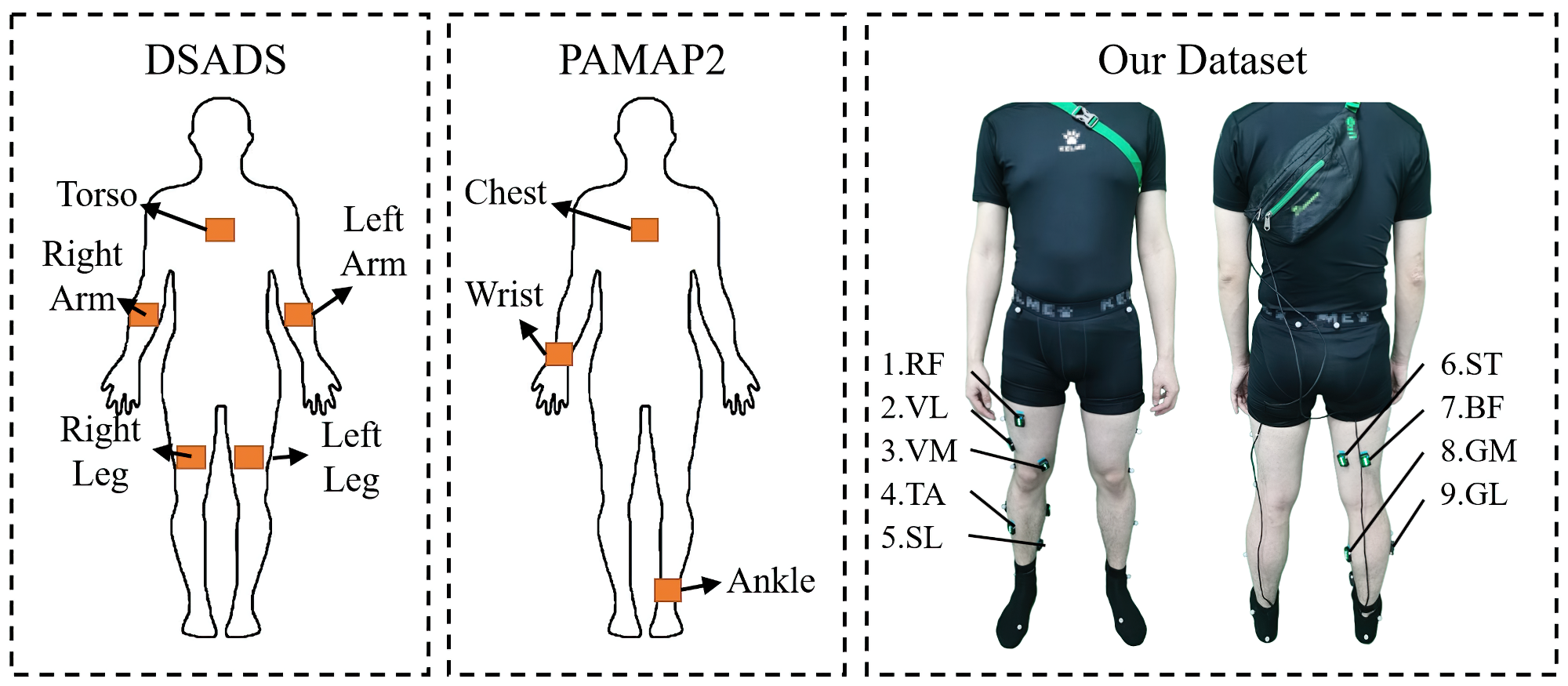}
  \caption{Location of the sensors in three datasets.}
  \label{dataset}
  \vspace{-6pt}
\end{figure}

Anther IMU-based dataset is the physical activity monitoring for aging people (PAMAP2) \cite{reiss2012introducing}. In the dataset, the activities are performed continuously, resulting in different time duration for each activity. Considering the time duration and occurred frequency in daily life, we choose 7 subjects and 5 activities: lying, standing, walking, cycling, and ascending stairs.

The EMG-based dataset was collected by our laboratory, which was approved by the Chinese Ethics Committee of Registering Clinical Trials. All the subjects signed the consent form before participating in the data collecting process, and they could decide to stop it at any time. The dataset contains ten subjects (age 25 $\pm$ 5 yr, height 1.70 $\pm$ 0.08m, weight 58.2 $\pm$ 10 kg) performing 5 activities: walking, ascending stairs, descending stairs, ascending ramp, descending ramp. Each activity lasts at least five minutes. Also, we label four gait phases based on the foot press sensors mounted on the heel and the first metatarsal bone, which form the second kind of label: heel strike, foot flat, heel off, toe off. The foot press signals are sampled at 500 Hz. 

Nine EMG sensors (Delsys Trigno, IM type \& Avanti type) are mounted on the subjects with a sampling frequency of 1111.11 Hz. The selected nine muscles are rectus femoris (RF), vastus lateralis (VL), vastus medialis (VM), tibialis anterior (TA), soleus (SL), biceps femoris (BF), semitendinosus (ST), gastrocnemius medial head (GM), and gastrocnemius lateral head (GL), which have been proven to be useful for HAR tasks \cite{naik2018ica}. 

\renewcommand\arraystretch{1.4}
\vspace{-6pt}
\begin{table}[bhtp]
\caption{An overview of the datasets used in this paper.}
\centering
\begin{tabular}{cccc}
\toprule
Dataset & Subjects & Activity Classes & Data Size of  Each Subject\\
\midrule
DSAD & 8 & 12&113040\\
PAMAP & 7 & 5&32615\\
OURS & 10 & 5 / 4 &25093\\
\bottomrule
\end{tabular}
\end{table}

\subsection{Experimental Setup}  
\subsubsection{Inter-method Evaluation} To evaluate the cross-user performance of our method via inter-method comparison, we select the representative methods that aim to solve the cross-subject issue. We introduce them briefly in the following. 

1. Unsupervised cross-subject adaptation based on maximizing classification discrepancy (MCD): MCD is a classic unsupervised domain adaptation framework in computer vision \cite{saito2018maximum}. The study of \cite{zhang2020unsupervised} leveraged this framework to solve the cross-subject problem and achieved state-of-the-art accuracy compared with advanced unsupervised methods.

2. Active learning for HAR (AL-HAR): The study of \cite{hossain2017active} implements the active-learning framework for sparse label-based HAR tasks. Also, they propose a dynamic k-means clustering method to improve the diversity and quantity of the core set. This method is implemented based on a machine learning model. In this paper, we implement this core set generation method to fine-tune the deep learning model.

3. Self-training for HAR (SelfHAR): The study of \cite{tang2021selfhar} proposed a self-training-based method that enables the model effectively learns to leverage a large unlabeled dataset to complement the small labeled target dataset.

4. Supervised fine-tuning: Supervised fine-tuning is a practical model transfer framework, where the parameters of the network are partially updated based on a middle-sized dataset. During model training, we fix the parameters of the CNN architecture and update the parameters of the fully-connected architecture.

We compare the above methods based on the three datasets with the same network architecture as our method. The hyperparameters of MCD, AL-HAR, and fine-tuning are the same as ours ( Section III-C-3). The learning rate of SelfHAR is 100 times that of our method since their framework enables the HAR model to be initially trained based on a small labeled dataset. The other hyper-parameters of SelfHAR are the same as ours. Additionally, we set the iteration number of L-HAR to be 3, which is the same as ours. 

\subsubsection{Intra-method Evaluation} From the description in Section III-B and Fig. 1, the ActiveSelfHAR framework contains four steps. Among these steps, the first and the last step are indispensable since the core set generation and the student model fine-tuning is essential for model transfer to the target domain. Thus, to evaluate the necessity of each step, we perform ablation experiments and form two frameworks: (1) unsupervised self-training framework (step 1 and 4), (2) unsupervised self-training with sparse labeling-based active learning to construct the sparse label-based semi-supervised learning framework (step 1, 2 and 4). These two frameworks and the SelfActiveHAR will be evaluated based on the three datasets. 

\subsubsection{Evaluation Metrics} Since we focus on the human activity pattern recognition task, the following statistical evaluation metrics have been adopted to quantify the recognition performance of our method: (1) precision, (2) recall, and (3) accuracy. The metrics are formulated as:
\begin{equation}
\label{eq_11}
\begin{split}
Precision &= \frac{TP}{TP+FP} \times 100\% \\
Recall &= \frac{TP}{TP+FN} \times 100\% \\
Accuracy &= \frac{TP+TN}{TP+TN+FP+FN} \times 100\% \\
\end{split}
\end{equation}
where TP, TN, FN, and FP denote the model output of true positive, true negative, false positive, and false negative, respectively. Aside from the aforementioned application-level metrics, two system-level indexes are adopted to comprehensively evaluate the model performance: (1) time cost: the time duration from data loading to model output to evaluate the computational complexity of the model, and (2) percentage of the labeled data: the percentage of the samples with actual labels in the dataset (including the test set) to quantify the labeling burden for model implementation.

\subsection{Experimental Results}  
\subsubsection{Cross-subject Performance Evaluation} This subsection examines our method by evaluating its application-level performance and system-level costs. Using such evaluations, we aim to demonstrate the feasibility of constructing such a framework and the potential of real-world applications. The results of our method depicted in all the figures below are yielded based on the iteration number of 3.

Fig. 4 depicts the performance comparison based on the DSAD dataset. The averaged cross-subject accuracy before the model transfer is 86.96$\%$. After model transfer based on the compared method, the average accuracies of MCD, AL-HAR, SelfHAR, ActiveSelfHAR, and fine-tuning are 89.21$\%$, 91.43$\%$, 93.69$\%$, 95.20$\%$, and 97.13$\%$, respectively. The results presented by our method are closest to the upper bound presented by supervised fine-tuning.
\begin{figure}[bhtp]
  \vspace{-6pt}
  \centering
  \includegraphics[width=\linewidth]{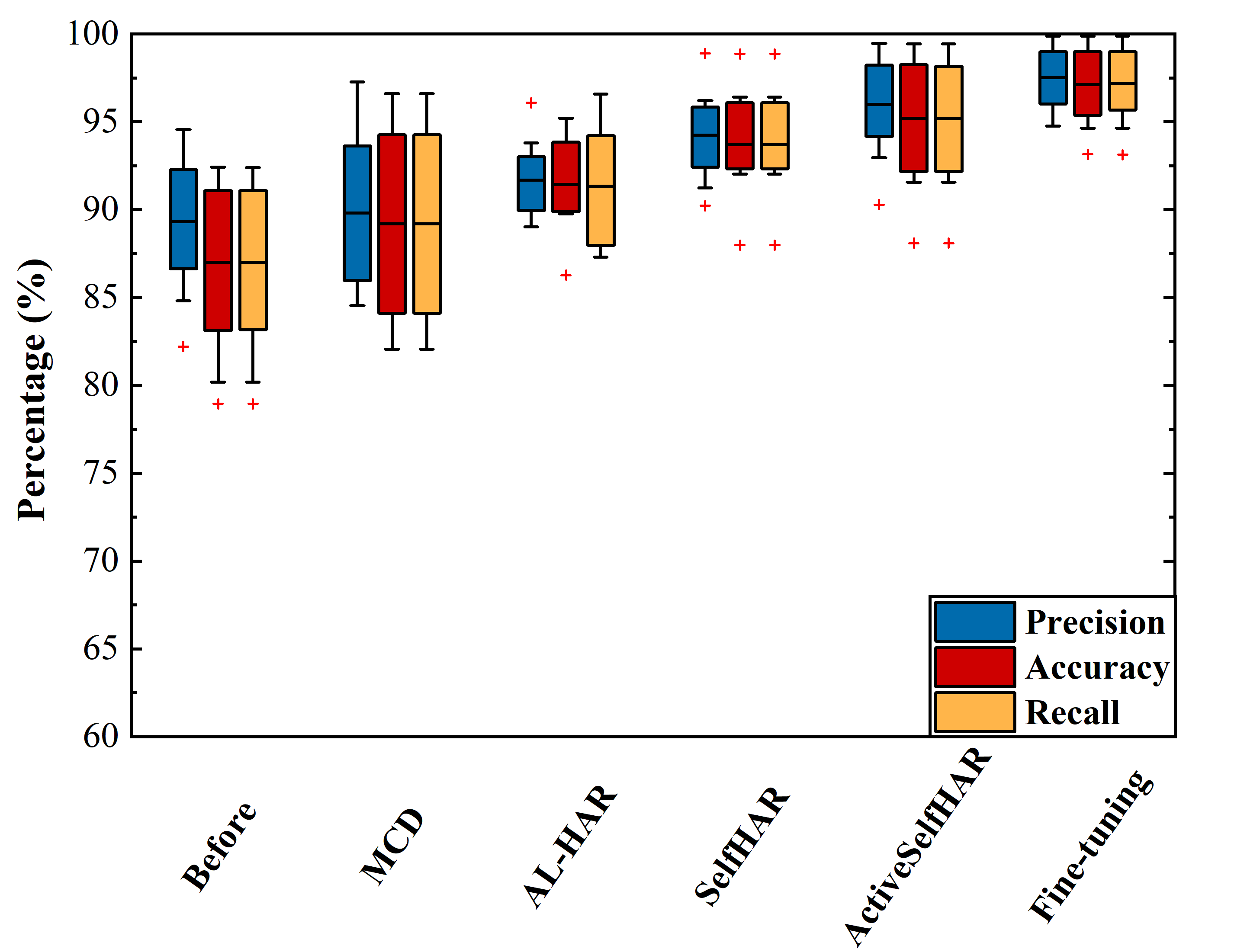}
  \caption{The precision, accuracy, and recall comparison on the DSAD datast.The horizontal line in the boxplot marks the median. The box ranges from the 25th to the 75th percentile.}
  \label{dataset}
  \vspace{-6pt}
\end{figure}

Table \ref{table-e1-DSAD} presents the system-level costs of each method on the DSAD dataset, i.e., the time cost and the percentage of the labeled data of the target dataset. Fine-tuning is the method that requires the most labeled data. Meanwhile, it presents the highest accuracy. AL-HAR is the most time-consuming method. Meanwhile, it requires the least labeled data. ActiveSelfHAR is the method that balances the time cost, the data labeling cost, and the model performance. 

\renewcommand\arraystretch{1.4}
\vspace{-6pt}
\begin{table}[bhtp]
\caption{The system-level costs and application-level performance conparison on DSAD dataset}
\centering
\begin{tabular}{cccc}
\toprule
Method & \makecell{Time Cost \\ (min)} & \makecell{Percentage of \\ the labeled data}  & Mean Accuracy ($\%$) \\
\midrule
MCD & 4.74 & N/A & 89.2\\
\makecell{AL-HAR \\ (iter = 3)} & 65.24 & 0.05 & 91.43 \\
SelfHAR & 3.25 & 10 & 93.69 \\
\makecell{ActiveSelfHAR \\ (iter = 1)} & 3.47 & 0.09 & 93.82 \\
\makecell{ActiveSelfHAR \\ (iter = 3)} & 8.92 & 0.17 & 95.20 \\
Fine-tuning & 1.33 & 43 & 97.13 \\
\bottomrule
\end{tabular}
\label{table-e1-DSAD}
\end{table}

Fig.5 depicts the mean accuracy comparison on the PAMAP dataset. Before model transfer, the averaged cross-subject accuracy is 49.58$\%$, suggesting that the data distribution is quite different between subjects. The average model transfer accuracies of MCD, AL-HAR, SelfHAR, ActiveSelfHAR, and fine-tuning are 80.02$\%$, 79.96$\%$, 77.07$\%$, 82.06$\%$, and 81.92$\%$, respectively. All the methods significantly increase the cross-subject model performance ($>$30$\%$). 
\begin{figure}[bhtp]
  \vspace{-6pt}
  \centering
  \includegraphics[width=\linewidth]{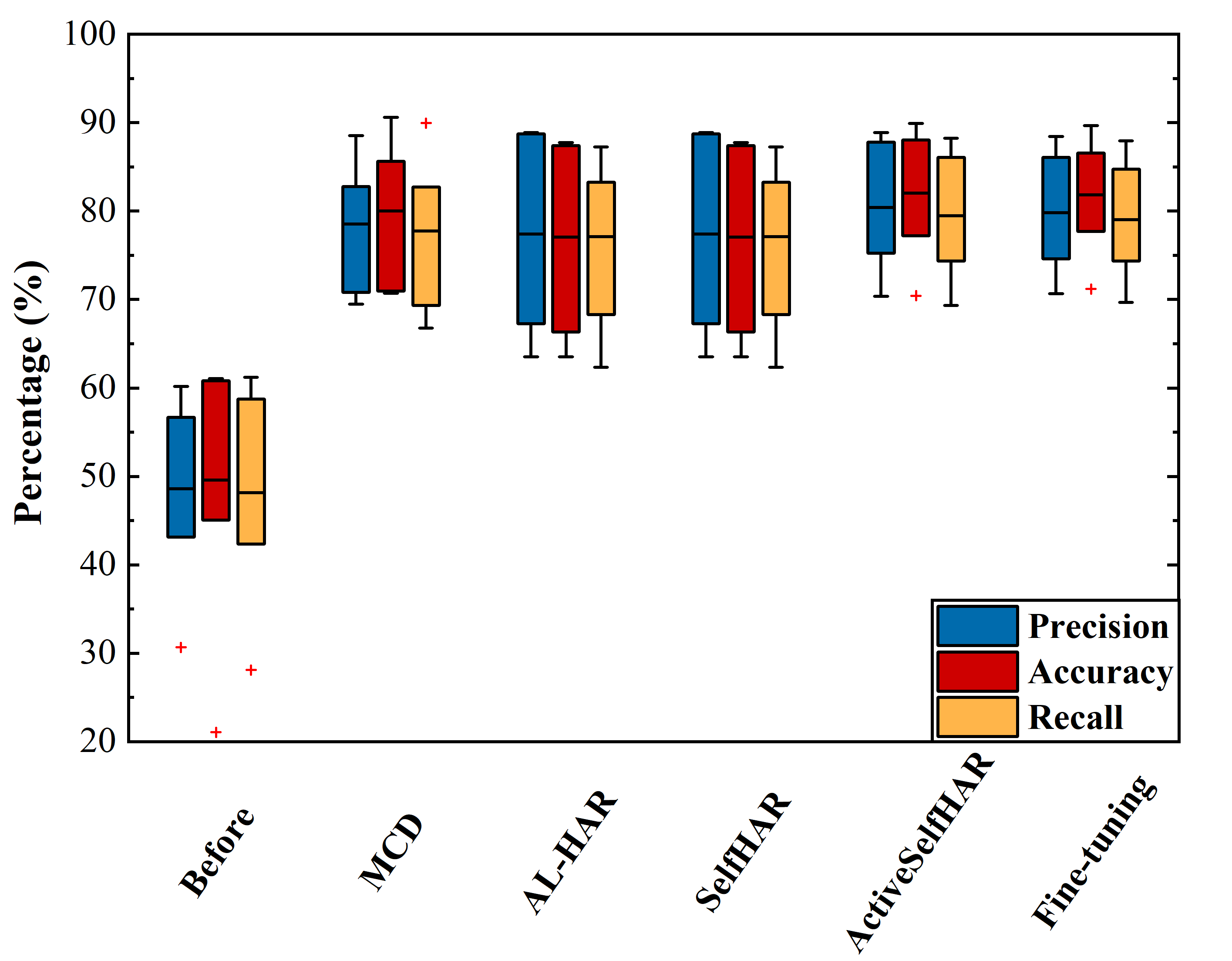}
  \caption{The precision, accuracy, and recall comparison on PAMAP dataset}
  \label{dataset}
  \vspace{-6pt}
\end{figure}

Among these methods, the mean accuracy of our semi-supervised-based method has reached the upper bound of the fully supervised fine-tuning (82.06$\%$ vs. 81.92$\%$). For each method, the variation of accuracies over all subjects is much larger than that on the DSAD dataset, which is depicted by the interquartile ranges of the boxes in the figure. This may be attributed to the cross-subject problem and the less data information provided by the PAMAP dataset due to the lower number of implemented wearable sensors

Table \ref{table-e1-PAMAP} presents each method's time cost and the degree of label requirement on the PAMAP dataset. ActiveSelfHAR presents equal accuracy with that of fine-tuning with significantly less labeled data. In addition, the cost time of our method with the maximum iteration number is more acceptable compared with the other two semi-supervised methods, i.e., AL-HAR and SelfHAR.

\renewcommand\arraystretch{1.4}
\vspace{-6pt}
\begin{table}[bhtp]
\caption{The system-level costs and application-level performance conparison on PAMAP dataset}
\centering
\begin{tabular}{cccc}
\toprule
Method & \makecell{Time Cost \\ (min)} & \makecell{Percentage of \\ the labeled data}  & Mean Accuracy ($\%$) \\
\midrule
MCD & 3.12 & N/A & 80.12\\
\makecell{AL-HAR \\ (iter = 3)} & 25.38 & 0.02 & 79.96 \\
SelfHAR & 7.85 & 10 & 77.07 \\
\makecell{ActiveSelfHAR \\ (iter = 1)} & 0.75 & 0.03 & 80.83\\
\makecell{ActiveSelfHAR \\ (iter = 3)} & 2.37 & 0.06 & 82.05\\
Fine-tuning & 0.43 & 43 & 81.92\\
\bottomrule
\end{tabular}
\label{table-e1-PAMAP}
\end{table}

The activity and phase recognition results of the methods on our dataset are depicted in Fig. 6(a) and Fig. 6(b), respectively. Before model transfer, the averaged cross-subject accuracy of activity and phase recognition is 70.94$\%$ and 79.62$\%$. The locomotion mode recognition performance of our method (91.81$\%$) is very close to that of fine-tuning framework (92.82$\%$). This compassion has verified the diversity of the core set generated by our method in the EMG-based dataset. The phase recognition results of our method (88.89$\%$) are a bit lower than those of SelfHAR (90.97$\%$) and fine-tuning (91.67$\%$). 

\renewcommand\arraystretch{1.4}
\vspace{-6pt}
\begin{table}[bhtp]
\caption{The system-level costs and application-level performance conparison on our dataset}
\centering
\begin{tabular}{cccc}
\toprule
Method & \makecell{Time Cost \\ (min)} & \makecell{Percentage of \\ the labeled data}  & \makecell{Mean Accuracy ($\%$) \\ (Activity/Phase)}\\
\midrule
MCD & 37.23 & N/A & 89.12 / 82.81\\
\makecell{AL-HAR \\ (iter = 3)} & 28.85 & 0.03 & 85.82 / 82.85\\
SelfHAR & 90.71 & 10 & 89.92 / 90.97 \\
\makecell{ActiveSelfHAR \\ (iter = 1)} & 4.17 & 0.21 & 90.26 /87.41 \\
\makecell{ActiveSelfHAR \\ (iter = 3)} & 13.18 & 0.75  & 91.79 /88.89 \\
Fine-tuning & 7.75 & 43  & 92.82 /91.67 \\
\bottomrule
\end{tabular}
\label{table-e1-Our}
\end{table}

Table \ref{table-e1-Our} presents the time cost and the percentage of the labeled data of each method based on our dataset. Since the data size is the same in locomotion and phase recognition tasks, the time cost and percentage of the labeled data are averaged over the locomotion and phase recognition tasks to form one table. The smaller batch size implemented on our dataset has dramatically increased the time cost of MCD and SelfHAR. For the other three methods, since no source data are imported, and no additional learning tasks are implemented in the model transfer process, the increments of time cost are relatively small.

\begin{figure}[h]
\vspace{-6pt}
\centering
\subfloat[]{\includegraphics[width=9cm]{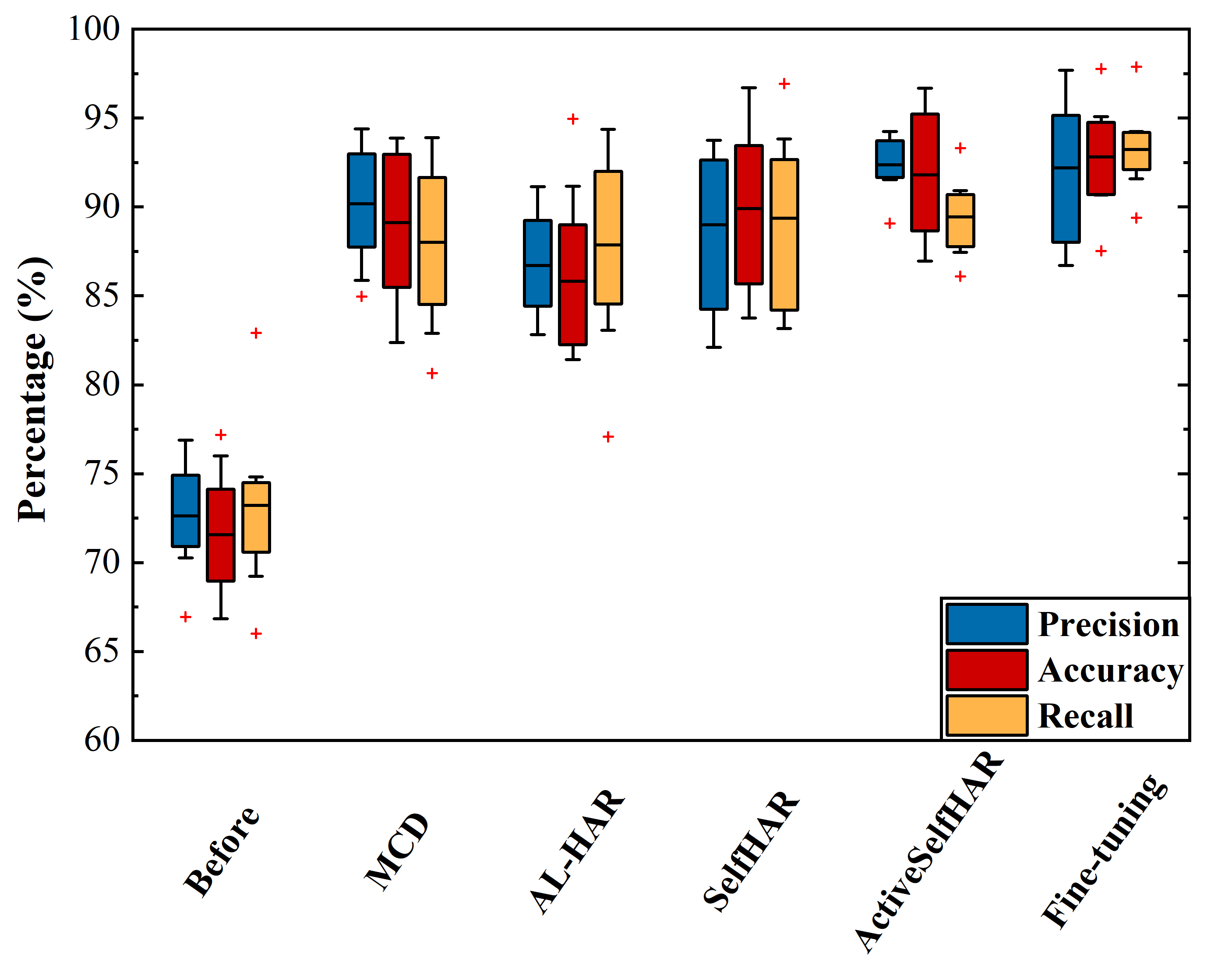}%
\label{fig6_01}}
\hfil
\subfloat[]{\includegraphics[width=9cm]{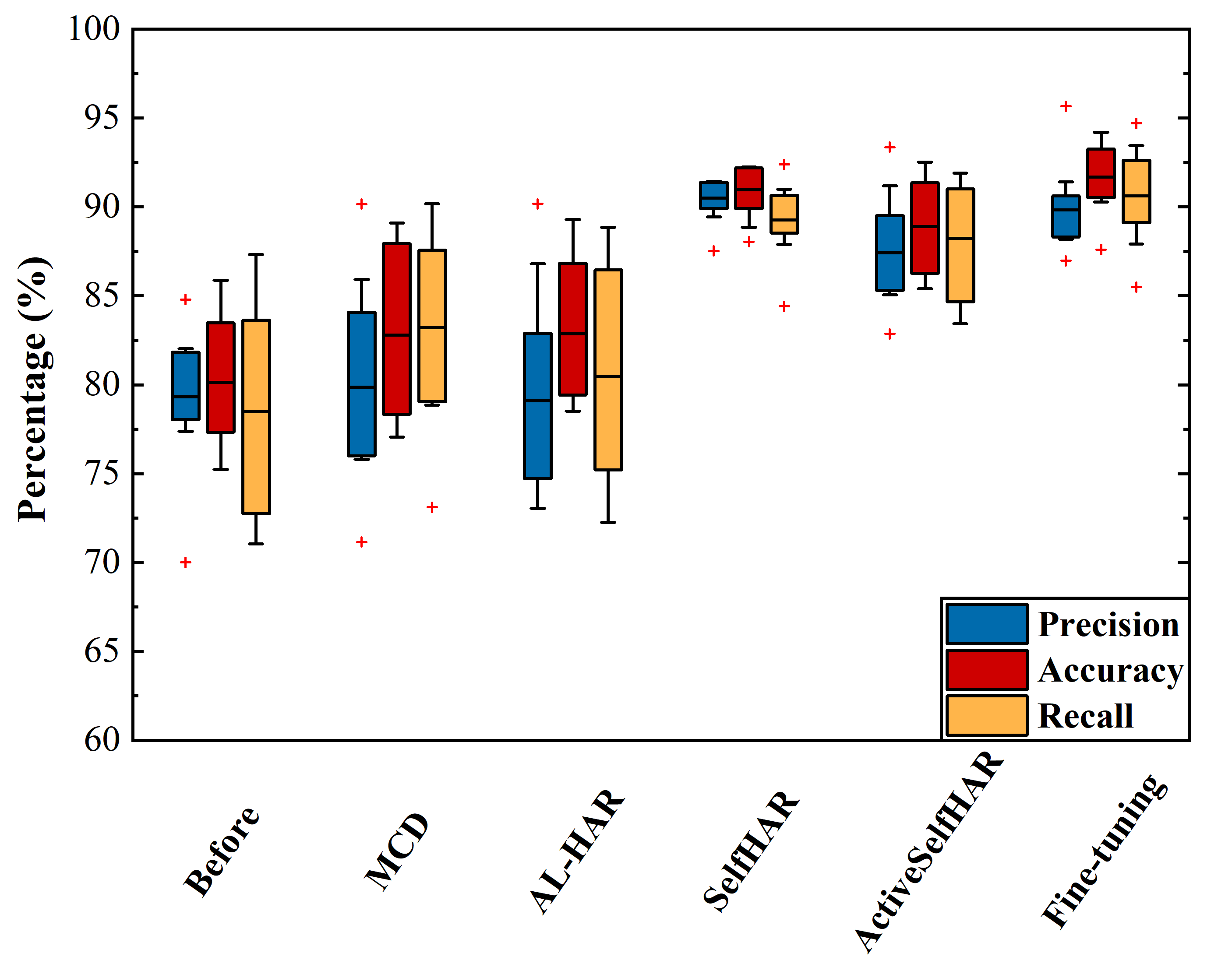}%
\label{fig6_02}}
\caption{The precision, accuracy, and recall comparison of (a) activity and (b) phase recogntion on our datast.}
\label{fig_6}
\vspace{-6pt}
\end{figure}

\subsubsection{Ablation Experiment} To evaluate the redundancy of the ActiveSelfHAR framework, we compare the contribution of each step in our framework to the performance and time cost. Based on the three datasets, we compare our framework with two sub-frameworks that are constructed based on different configurations: (1) unsupervised self-training framework (Sub-UST) and (2) sparse-label-based semi-supervised framework (Sub-SLSS). The detailed description of these two frameworks is in Section IV-B-2. Additionally, only the results with the iteration number of 3 are presented since there is no big difference in time cost and model performance between the results of 1 and 3.

We first evaluate the performances of the frameworks based on the DSAD and PAMAP datasets, shown in Table \ref{table-e2-DSAD-PAMAP}. For performance comparison on the DSDA dataset, the unsupervised framework of Sub-UST has increased the performance from 86.96$\%$ to 89.14$\%$. This result has proved the effectiveness of the self-training framework. The sparsely labeled sample provided by Sub-SLSS has little impact on the performance improvement (89.14$\%$ to 89.71$\%$). After generating the informative set based on the discrete labeled samples and using it to construct the core set with the self-training set (step 3), the performance is significantly increased (89.71$\%$ to 95.20$\%$) while the standard deviation is decreased (6.45$\%$ to 3.98$\%$). 

For the PAMAP dataset, the Sub-UST framework has significantly increased the cross-subject performance (49.58$\%$ to 80.84$\%$). The sparse labeled samples provided by Sub-SLSS have little effect on performance improvement. The informative set generation process of ActiveSelfHAR does not work well in the PAMAP dataset. This may be attributed to the error accumulated from the process of self-training set construction, causing the locations of the class centers that are essential for informative sample selection to be far from the actual centroid of the classes. Based on the two datasets, the time cost comparison of the three frameworks suggests that the most time-consuming step in our framework is step 3, i.e., the core set augmentation process. 

\renewcommand\arraystretch{1.4}
\vspace{-6pt}
\begin{table}[bhtp]
\caption{The system-level costs and application-level performance conparison of the three frameworks on DSAD and PAMAP dataset}
\centering
\begin{tabular}{ccc}
\toprule
Method & \makecell{Time Cost (min)\\(DSAD/PAMAP)} &  \makecell{Mean (Standard Deviation) ($\%$) \\ (DSAD/PAMAP)}\\
\midrule
Sub-UST & 3.45 / 1.29 &  89.12(6.56) / 80.84(6.31)\\
Sub-SLSS & 4.09 / 1.43 &  89.71(6.45) / 81.18(6.16)\\
ActiveSelfHAR & 8.92 / 2.37 &  95.20(3.98) / 82.05(5.74)\\
\bottomrule
\end{tabular}
\label{table-e2-DSAD-PAMAP}
\end{table}

The results of the three frameworks on our dataset are shown in Table \ref{table-e2-Our}. The Sub-UST framework has improved the average cross-subject accuracy for activity and phase recognition (activity: 70.94$\%$ to 86.49$\%$, phase: 79.62$\%$ to 82.43$\%$). Similar to the performance on IMU-based datasets, the improvement effect of Sub-SLSS on the EMG-based activity and phase recognition accuracy is not significant. The core set construction process of ActiveSelfHAR has significantly increased the activity and phase recognition performances (activity: 84.12$\%$ to 88.89$\%$, phase: 87.59$\%$ to 91.79$\%$). The time cost of the three frameworks on our dataset suggests that the core set construction process is more time-consuming than other steps. 

The results of the three datasets suggest that except for the indispensable model fine-tuning process (step 4), the self-training set generation (step 1) and the core set generation process based on the informative samples (step 2 and 3) have significantly improved the cross-subject performance of the model, while the informative sample selection alone (step 2) have little contribution to the improvement of model performance. For the time cost of the three frameworks on the three datasets, the self-training framework (step 1 and 4) accounts for nearly half of the total time cost, while most of the cost is spent on the indispensable model fine-tuning phase. The core set construction process (step 3) occupies another half of the time cost. Because of the increased average accuracy and the reduced variance, the time cost is acceptable. 

\renewcommand\arraystretch{1.4}
\vspace{-6pt}
\begin{table}[bhtp]
\caption{The system-level costs and application-level performance conparison of the three frameworks on our dataset}
\centering
\begin{tabular}{ccc}
\toprule
Method & \makecell{Time Cost (min)\\(Activity/Phase)} &  \makecell{Mean (Standard Deviation) ($\%$) \\ (Activity/Phase)}\\
\midrule
Sub-UST & 8.18 / 6.21 &  86.49(5.14) / 82.43(4.26)\\
Sub-SLSS & 8.52 / 6.43 &  87.59(5.06) / 84.12(4.14)\\
ActiveSelfHAR & 8.92 / 11.37 &  91.79(3.58) / 88.89(2.65)\\
\bottomrule
\end{tabular}
\label{table-e2-Our}
\end{table}

\section{Discussion}
In this paper, we propose a semi-supervised training framework that incorporates the self-training framework into an active learning strategy to enable the HAR model to adapt to a new user based on sparsely labeled data. Through the method, We aim to construct a diverse core set based on the sparsely labeled sample to satisfy the data requirement of the deep network training process and enable the learned network to be competitive over the target domain. We demonstrate our contributions by comparing the methods akin to ours and evaluating our methods on IMU-based and EMG-based datasets. Additionally, we evaluate the redundancy of the framework by comparing the time cost of each step and the contribution of each step to the improvement of the cross-subject performance of the model. The results demonstrate our method's feasibility.

\textbf{Performance of cross-subject HAR methods:} Based on the three datasets, our method achieves the best all-around performance in three aspects: cross-subject classification accuracy, time cost, and data labeling cost. Furthermore, if relaxing the constraints of time and data labeling cost, appropriately increasing the number of iterations and queried samples would enable the performance of our method to approximate that of fully supervised learning.

By comparing the performance of MCD with those of other methods, it suggests that MCD is less stable and accurate, which may be attributed to the easily biased supervision information extracted from domain intrinsic structures. For the related semi-supervised methods, i.e., SelfHAR and AL-HAR, their performance is less satisfactory on the PAMAP dataset. SelfHAR initializes the deep model based on a small labeled dataset. This process would lead to model overfitting or underfitting problem, making it challenging to tune the hyper-parameters of the network. The significant difference in data distribution of subjects may induce more uncertainty in this process, leading to a large variance in the results of different subjects (Fig. 6(b)). AL-HAR is also easily affected by the data quality of the dataset since the label of each cluster is aligned with that of the most informative sample in that cluster. If a cluster contains samples of multi classes, this process would induce errors in the labels. This may be the reason for the lower mean accuracy of AL-HAR on the PAMAP dataset. Although fine-tuning achieves the highest mean accuracies on the three datasets, it is relatively difficult to be implemented in real-world scenarios because of the high labeling cost. 

A noteworthy phenomenon is that the phase recognition results of our method are lower than those of SelfHAR and fine-tuning (Fig. 6(b)). The possible reason is as follows. Compared with activity labels, the transitions between phase labels are much more frequent, enabling more samples in the class boundaries. Different from the batch-level data labeling strategy of SelfHAR and fine-tuning, the discrete informative samples are selected to label in our framework. The samples of the resulting informative set and the self-training set are less informative than the labeled samples. Thus, it is difficult for the model to learn the entire distributions of the boundaries based on our method since the samples closest to the boundaries are likely to be as informative as the samples we select to label. It can be further improved by incorporating batch-labeling-based few-shot learning methods in the framework.

\textbf{Costs of cross-subject HAR methods:} For the system-level cost, the data labeling cost of our method is much smaller than fine-tuning and SelfHAR. Despite the lower cost of AL-HAR and MCD, our data labeling cost is acceptable, considering the accuracy and stability of the model performance. As to the time cost, our method exhibits a moderate time cost compared with other methods. Furthermore, the setting of ActiveSelfHAR can be changed according to the system constraint. If the time cost is limited, setting the number of iterations to 1 still enables the model to present an acceptable accuracy.

Additionally, since the model training phases of SelfHAR and MCD require source data and additional learning tasks, their time cost is easily affected by the batch size (Table \ref{table-e1-Our}). This constraint would not have a significant impact on our method. AL-HAR has a relatively high time cost since each sample needs to calculate its silhouette coefficient and information entropy to measure the information it carries. In our framework, the information carried by each sample is measured based on the distance to the sparse class centers, which makes our method more time-efficient.

\textbf{Necessity of each step in ActiveSelfHAR:} By comparing the performances of Sub-UST and Sub-SLSS with that of the whole framework on the three datasets, it suggests that the group-level data labeling process can significantly improve the performance of the model (step 1 and step 3). On the one hand, the augmented core set generated by the sparse labeled samples and spatial-temporal constraints provides reliable information to enable the model to mitigate the biasness of distributions of source and target domains. The well-trained model would in turn iteratively increase the reliability of the self-training set constructed by the self-training framework. On the other hand, group-level labeled data is likely to have a more statistically significant impact on the model performance compared with sparse sample points due to the local optimization methods of deep model training \cite{sener2017active}. Except for the commonly used model fine-tuning process in related methods (step 4), step 2 only has little impact on the model performance. We hardly consider this step redundant since it is indispensable for the following augmented core set construction process. As to the time cost, the augmented core set construction process is the most time-consuming, which selects similar samples by traversing the entire dataset for each informative sample. Considering its contribution to the model performance, this step should be further optimized rather than removed. When selecting similar samples for each sample in the core set, the time cost can be reduced by leveraging clustering or spatial similarity methods to predefine a small data group instead of the entire dataset. 

Another limitation is the generation of the accumulated error, which is shown in the results of PAMAP datasets (Table \ref{table-e2-DSAD-PAMAP}). The significant difference in data distribution of the subjects (cross-subject accuracy of 49.58$\%$) would induce significant uncertainty in the self-training set generation process, causing the deviation between the calculated class centers and the actual class centroids. This error would  accumulate along the pipeline of our framework, propagating to the selected informative and augmented samples. Although the accumulated error does not have a negative impact on the self-training framework, it makes the active-learning framework (step 2 and 3) redundant in the pipeline. This issue can be addressed by adding supervision information to the core set construction and augmentation process.
  
\section{Conclusion}
Aiming to overcome the obstacle of cross-subject issue to the real-world implementation of HAR methods, we introduce the ActiveSelfHAR: an iterative training framework that complements active learning with self-training. Our method utilizes the confidence output of the self-training method to effectively augment the core set of active learning and enable the learned HAR model to be competitive over the target domain, i.e., the new users. We use two public IMU-based public datasets and construct an EMG-based dataset to evaluate the versatility of our method and compare various related methods to evaluate the effectiveness of our method. The extensive experiments demonstrate that the presented method presents similar results to the upper bound, i.e., the supervised fine-tuning framework with significantly less data labeling cost (less than 1$\%$ labeled data of the target dataset). The all-around performance of our method outperforms other unsupervised or semi-supervised frameworks for similar tasks. The ablation experiment demonstrates the necessity of each step in the overall cross-subject performance. This method presents a promising capability of using the incorporated framework to address the cross-user issue, which will aid the further implementation of user-independent HAR methods into smart home healthcare systems. 

\bibliographystyle{IEEEtran}
\bibliography{IEEEabrv,mycite}

\end{document}